\documentclass[]{spie}  

\usepackage{amsmath,amsfonts,amssymb}
\usepackage{graphicx}
\usepackage{hyperref}
\hypersetup{colorlinks=true, linkcolor=blue, allcolors=blue}
\usepackage{caption}
\usepackage{subcaption}
\usepackage{graphicx}

\title{The XRISM Pipeline Software System: Connecting Continents, Processes, Testing, and Scientists}

\author[1,2]{Trisha F. Doyle}
\author[2]{Matthew P. Holland}
\author[2,3]{Robert S. Hill}
\author[2,5,6]{Tahir Yaqoob}
\author[2,5,7]{Mike Loewenstein}
\author[4]{Eric D. Miller}
\author[1,2]{Patricia L. Hall}
\author[2,3]{Efrem Braun}
\author[2,3]{Efrain Perez-Solis}
\affil[1]{Innovim, LLC, 6401 Golden Triangle Dr, Greenbelt, MD, USA}
\affil[2]{NASA, Goddard Space Flight Center, Greenbelt, Maryland, USA}
\affil[3]{ADNET Systems, Inc., 6720B Rockledge Drive, Bethesda, MD, USA}
\affil[4]{Massachusetts Institute of Technology, Kavli Institute for Astrophysics and Space Research, Cambridge, MA, USA}
\affil[5]{Center for Research and Exploration in Space Science and Technology (CRESST), NASA/GSFC, Greenbelt, MD USA}
\affil[6]{University of Maryland Baltimore County, Center for Space Sciences and Technology, Baltimore, MD, USA}
\affil[7]{University of Maryland, Department of Astronomy, College Park, MD, USA}

\authorinfo{Further author information: (Send correspondence to T.F.D)\\T.F.D.: E-mail: trisha.f.doyle@nasa.gov}

\begin{document} 
\maketitle

\begin{abstract}
XRISM (X-Ray Imaging and Spectroscopy Mission), with the Resolve high-resolution spectrometer and the Xtend wide-field imager on-board, is designed to build on the successes of the abbreviated Hitomi mission to address outstanding astrophysical questions using high resolution X-ray spectroscopy. In preparation for launch, the XRISM Science Data Center (SDC) is constructing and testing an integrated and automated system for data transfer and processing based upon the Hitomi framework, introducing improvements informed by previous experience and internal collaboration. The XRISM pipeline ingests FITS files transferred from Japan that contain data converted from spacecraft telemetry, processes (calibrates and screens) the data, creates data products, and transfers data and metadata used to populate data archives in  the U.S. and Japan. Improvement and rigorous testing of the system are conducted from the single-task level through fully-integrated levels. We provide an overview of the XRISM pipeline system, with a focus on the data processing, and how new and improved documentation and testing are creating accessible and effective software tools for future XRISM data.
\end{abstract}


\section{INTRODUCTION}
\label{sec:intro} 
The X-Ray Imaging and Spectroscopy Mission (XRISM) is the next X-ray observatory to be launched no earlier than Japanese fiscal year 2022.  This mission is a joint operation between the Japanese Aerospace Exploration Agency (JAXA), the National Aeronautical and Space Administration (NASA), and the European Space Agency (ESA).  The XRISM mission is a re-fly of a previous mission, Hitomi, which met an unfortunate and early end to its tenure.  Before Hitomi was lost however, unprecedented high-resolution X-ray observations were obtained during commissioning, notably those of the Perseus galaxy cluster, which fueled the motion to re-fly the mission and recapture the science.  The XRISM satellite will be flown with two (out of the original four) instruments, the Resolve spectrograph (micro-calorimeter) and Xtend wide field imager, which is a charge coupled device (CCD), both of which are sensitive in soft X-ray wavelengths; the high energy instruments from Hitomi were scrubbed for the XRISM mission.

As part of the larger XRISM mission team in the US, the XRISM Science Data Center (SDC) is responsible for the main processing pipeline, reprocessing tools, and quick-look analysis tools for processing the science and calibration data.  We work with both Resolve (US and Japan) and Xtend (Japan) instrument teams to obtain the latest calibration data to improve our existing tools and pipeline, as well as the calibration database (CalDB).   Although much of the software is reused from Hitomi, updates have resulted from instrument changes or new calibration information. A more rigorous testing program has uncovered a few bugs. Individual programs have been refactored to accommodate these changes, along with the routine adaptations required to shift to a new mission. We have used lessons learned from Hitomi, refined our software engineering practices, and taken advantage of the additional time afforded by supporting a new mission. The result is an efficient and well-documented pipeline for processing XRISM science and calibration data.

We discuss in this manuscript an overview of the full XRISM pipeline in \S\ref{sec:pipeline}, the data sets used to test the pipeline and individual tasks in \S\ref{sec:minimission}, the improved framework the XRISM SDC has set up to test the individual tasks in \S\ref{sec:unittests}, the improved and new documentation of both code and user help in \S\ref{sec:documentation}, and finally the improvements we have made since Hitomi to make XRISM even more successful in \S\ref{sec:improvements}.

\section{The XRISM Pipeline}
\label{sec:pipeline}

The XRISM pipeline is a daemon-driven collection of scripts that processes the data delivered from the pre-pipeline (PPL) from Japan and produces a data set, associated with an observation (henceforth called a sequence), that is placed in data archives and available to scientists.  The  pre-pipeline \cite{Eguchi2022} (see Eguchi et al., these proceedings) obtains telemetry data from the spacecraft and converts this data into first FITS files (FFFs), which are then transmitted to the pipeline (PL) in the US.  The XRISM SDC and PL are responsible for ingesting and processing the FFFs, and archiving and encrypting the processed data.  The XRISM pipeline processing center is located on the Goddard Space Flight Center (GSFC) campus of NASA in Greenbelt, MD, USA. The pre-pipeline processing center is located at the Institute of Space and Aeronautical Science (ISAS) on the Sagamihara campus of JAXA in the Kanagawa prefecture in Japan.

The XRISM pipeline is executed on a series of Linux virtual machines (VMs), which includes a small network of pipeline machines: five processing machines, which can run the processing script (the core pipeline processing of data) in five separate instances, and the main processing machine, which runs all the other daemons not associated with the processing script, including fetching and archiving the data.  Parallelization is achieved by processing a single and distinct sequence on each of the five processing machines simultaneously.  

The XRISM pipeline is a collection of Perl modules linked together and executed by a series of daemons.  The pipeline is triggered when the \emph{pre-fetch} daemon, which is a cron job, notices that data is ready to be transmitted from the PPL.  The daemons are always running in the background, unless stopped by a pipeline operator, so once the \emph{pre-fetch} cron job is invoked, the pipeline runs autonomously through the following daemons, described in more detail below: \emph{fetch}, \emph{stream}, \emph{post\_proc}, \emph{pre\_archive}, \emph{archive}, and \emph{deleteit}.

The high-level overview of the pipeline processing steps from the receiving of FFF data from the PPL to the distribution of processed data to the archives is as follows:
\begin{itemize}
    \item \emph{pre-fetch} daemon (cron job): XRISM version of the data transfer system (xDTS) is executed to transfer data from the ISAS PPL to the GSFC PL
    \item \emph{pre-fetch} daemon: data archive system (DAS) invoked to copy the FFFs to the deep archive
    \item \emph{fetch} daemon: data is verified in the pre-processing (pre-proc) area, and prepared for processing through the pipeline script
    \item \emph{stream} daemon: data is processed through the pipeline script in the processing (proc) area
    \item \emph{post\_proc} daemon: validate processed data and check for errors from the pipeline processing script in the post-processing (post-proc) area
    \item \emph{pre\_archive} daemon: build sub-directories for archive, compress and encrypt necessary data
    \item \emph{archive} daemon: move processed data to local archive (deep archive) and remote archive hosted at GSFC
    \item \emph{archive} daemon: xDTS is executed again to move processed and encrypted data from the PL back to the ISAS archive
    \item \emph{archive} daemon: once archived, data can be provided to scientists based on PI (proprietary), or public availability
    \item \emph{deleteit} daemon: cleans up post-processing area
\end{itemize}

DTS and DAS are inherited from previous high-energy missions.  xDTS is the XRISM-adapted version of DTS, which is used in the current XRISM pipeline instead of the canonical DTS, to account for updates and changes in communication channels between ISAS and GSFC.  
The software tasks used in the main part of the processing pipeline (by means of the \emph{stream} daemon) are part of a larger suite of software tools used for general FITS file manipulation, 
known as FTOOLS \cite{ftoolsweb,Blackburn1995}.  The FTOOLS software package is part of the broader High Energy Astrophysics Software (HEASoft \cite{heasoft,Blackburn1995}) suite, which is hosted and maintained by the High Energy Astrophysics Science Archive Research Center (HEASARC \cite{heasarc}) at NASA/GSFC.  The HEASARC is also where archived XRISM data is hosted in the US.  HEASoft is a collection of FTOOLS and high-energy mission-specific software tasks including software for both legacy and current missions, such as Suzaku, NICER, Swift, and IXPE.  The software tasks used in the processing pipeline are both the general FTOOLS and XRISM mission-specific tasks.  The XRISM mission-specific tasks will be incorporated into the HEASoft suite near or soon after launch.  These mission-specific tasks are used for processing and re-processing (by a user) and include the following key functionalities, among others: converting pulse height amplitude (PHA) to pulse invariant (PI), creating response matrix functions (RMFs), secondary corrections, modulated X-ray source (MXS) operations, and event flagging.  For more specifics on some of the SDC software tasks, see the SDC SPIE 2020 paper \cite{loewenstein2020}.  These tasks use information from auxiliary and mission-dependent files, calibration files, and housekeeping files to perform their prescribed function.  The XRISM SDC is responsible for 62 tasks, 15 under the general HEASoft umbrella and 47 under the XRISM-specific tasks.  Under the XRISM-specific tasks, we maintain 16 Resolve tasks, 5 Xtend tasks, 10 general tasks, and 16 multi-instrument tasks.  Additionally, the XRISM SDC is responsible for maintaining 16 libraries, 4 in the general HEASoft area and 12 in the XRISM-specific area.  These libraries are used by multiple tasks throughout the XRISM software.

\begin{figure}[hbt]
    \centering
    \includegraphics[width=0.65 \textwidth]{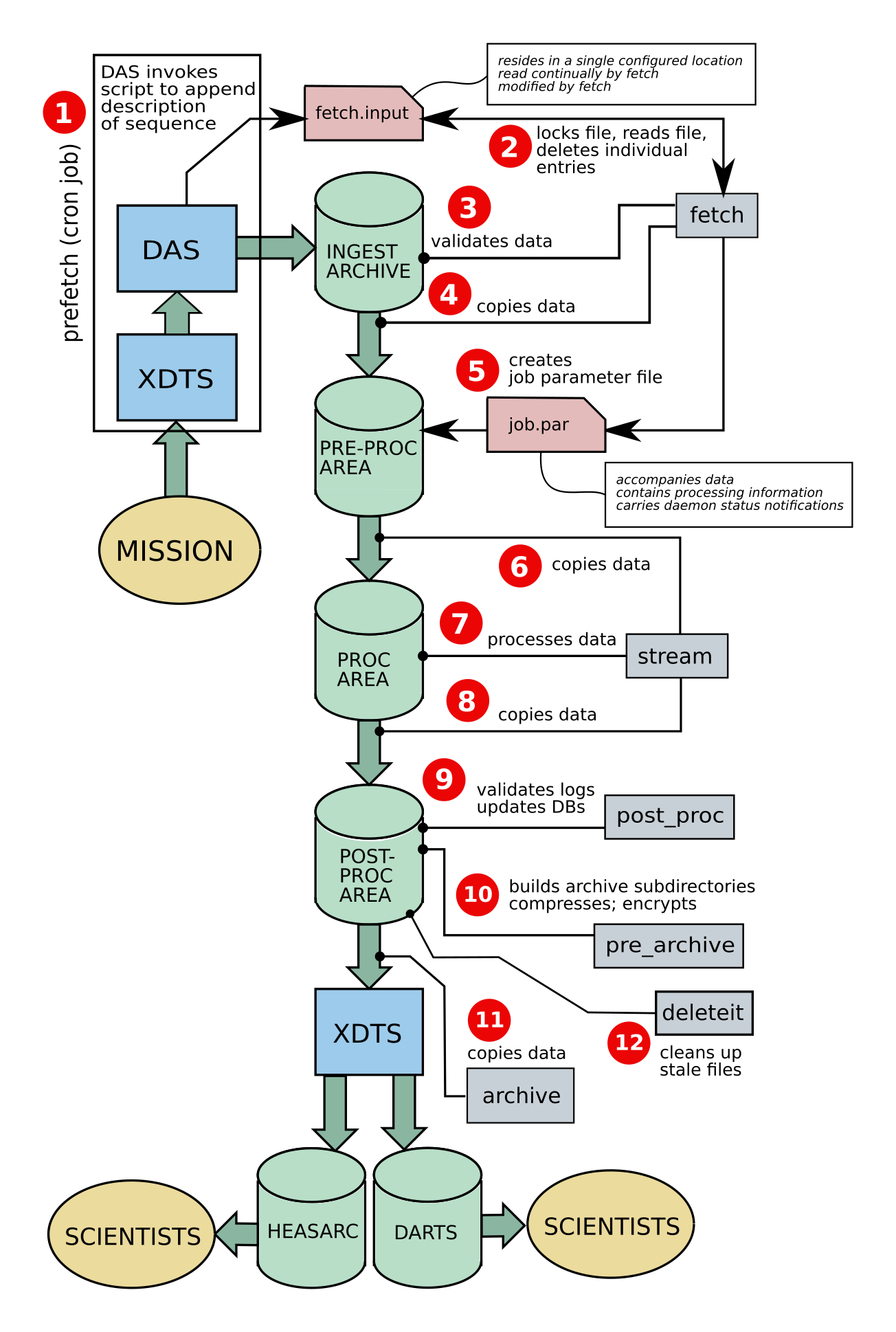}
    \caption{Data flow and processing steps applied by the SDC starting with the acquisition of FFF, and concluding with data delivery to scientific investigators. Intermediate steps where the data is copied for backup storage are also indicated, but not discussed further.}
    \label{fig:plflow}
\end{figure}

We discuss here the more detailed flow of pipeline processes, which can also be seen in the flow chart in Fig.~\ref{fig:plflow}.  The pipeline is always ``on'' and waiting, as it checks periodically for data transmissions.  The \emph{pre-fetch} daemon, which is the start, or trigger, of the pipeline, is a cron job that waits for input data.  The pipeline is triggered by a data delivery from the PPL, which is checked for by the \emph{pre-fetch} daemon.  Once there is data ready to be processed,  data is transferred from JAXA/ISAS, after processing through the PPL, via xDTS to GSFC.  DAS is then invoked and notifies GSFC that there is data ready for transmission to and processing through the pipeline, while simultaneously copying the data to the local deep archive at GSFC, where a copy of the PPL output data is stored.  Additionally, in the pre-fetch stage, the databases associated with the observation are also transferred from Japan via xDTS.  The \emph{fetch} daemon then validates and copies the data to the pre-proc area.  The \emph{fetch} daemon also validates and copies the databases to the ingest archive and installs them in the working directory.  Next, the \emph{stream} daemon is activated.  The \emph{stream} daemon first copies the data from the pre-proc area to the proc area.  The \emph{stream} daemon executes the pipeline processing script in the proc area, which is the script that executes the series of FTOOLS and XRISM-specific tasks that perform the calibration on, and process the data.  The \emph{stream} daemon processes a single sequence, associated with an observation.  However, the \emph{stream} daemon can be run on separate machines to process different sequences simultaneously.  After the sequence has been processed by the pipeline script, the \emph{stream} daemon checks for severe errors, copies the output to the post-proc area (if no severe errors have occurred), and cleans up the processing area.  The \emph{post\_proc} daemon then validates the processed data in the post-proc area and updates the databases accordingly.  The \emph{pre\_archive} daemon then creates the sub-directory structure for archiving the data, compresses and encrypts data as configured, and updates the catalog file.  The \emph{archive} daemon then copies the compressed data into the local archive and invokes DTS to initiate the transfer of data to the JAXA/ISAS Data ARchives and Transmission System  (DARTS \cite{darts}) and HEASARC at NASA/GSFC.  Finally, the \emph{deleteit} daemon deletes the data from the post-proc area and cleans up any other extraneous files.

\begin{figure}[h]
     \centering
     \begin{subfigure}[b]{0.3\textwidth}
         \centering
         \includegraphics[width=\textwidth]{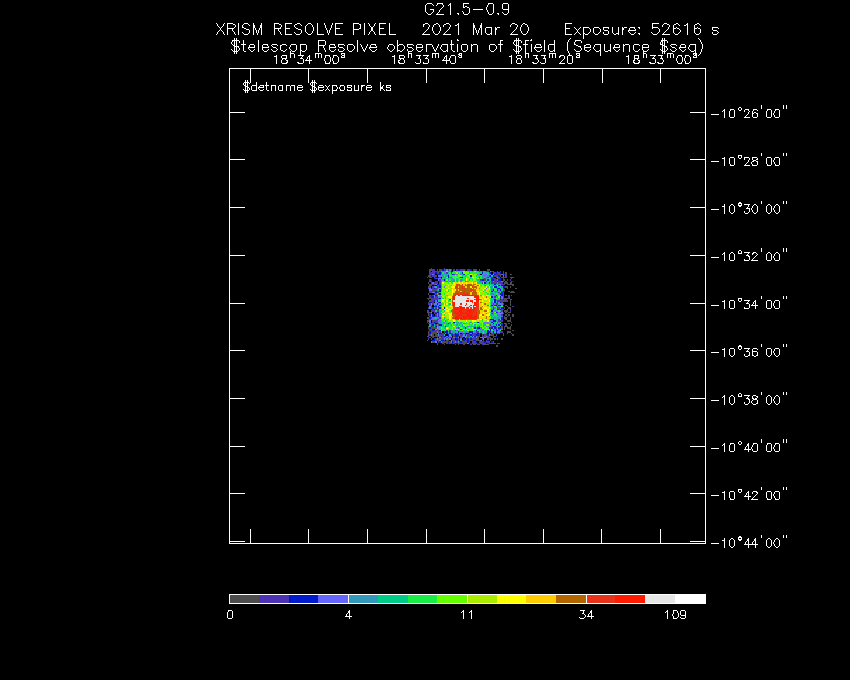}
         \caption{XRISM/Resolve image}
         \label{fig:rslimg}
     \end{subfigure}
     \hfill
     \begin{subfigure}[b]{0.3\textwidth}
         \centering
         \includegraphics[width=\textwidth]{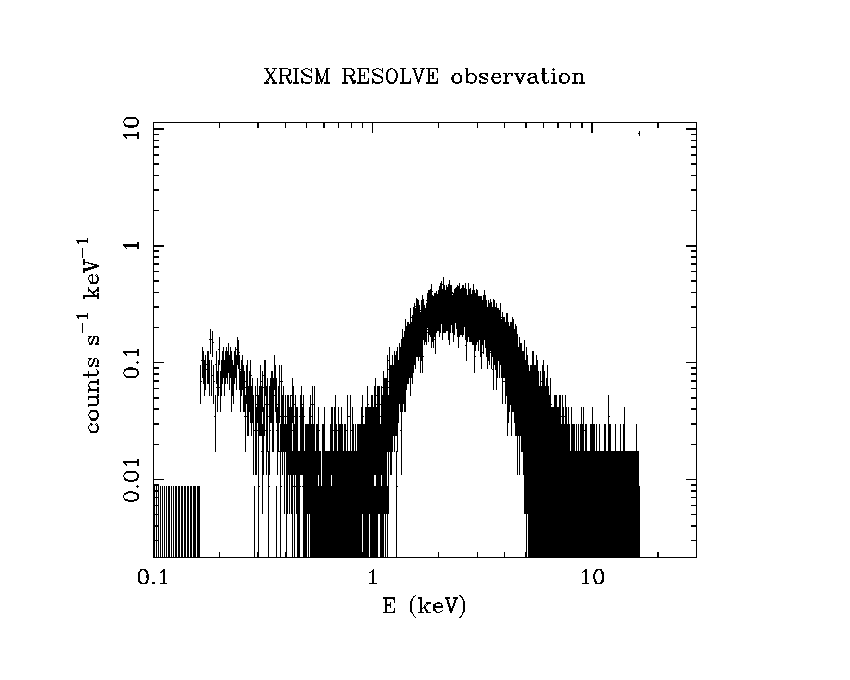}
         \caption{XRISM/Resolve spectrum}
         \label{fig:rslspec}
     \end{subfigure}
     \hfill
     \begin{subfigure}[b]{0.3\textwidth}
         \centering
         \includegraphics[width=\textwidth]{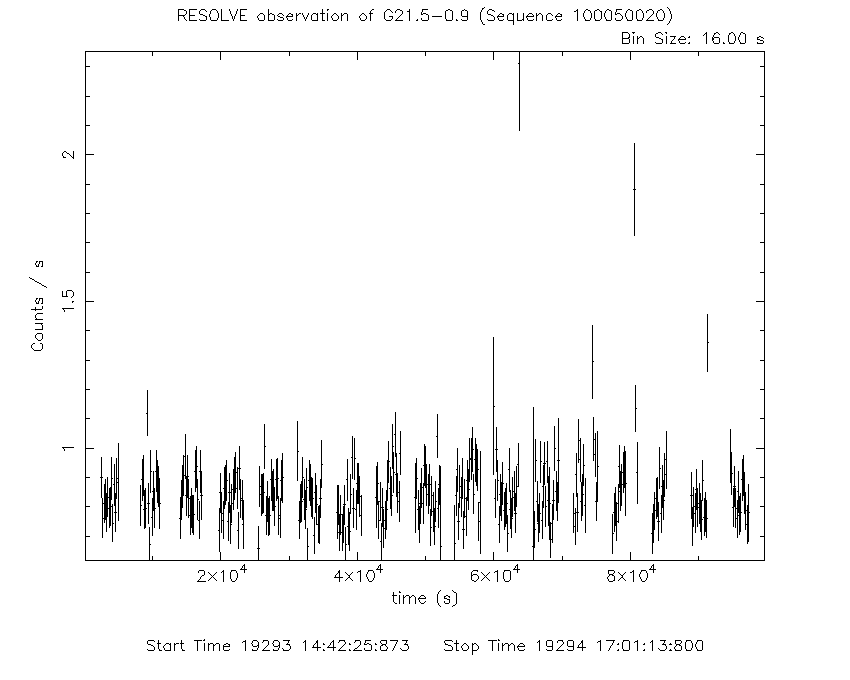}
         \caption{XRISM/Resolve light curve}
         \label{fig:rsllc}
     \end{subfigure}
        \caption{XRISM/Resolve pipeline products from the the Hitomi observation 100050020, converted to be compatible with XRISM software, see \S\ref{sec:minimission} for more details.}
        \label{fig:prodrsl}
\end{figure}

\begin{figure}[h]
     \centering
     \begin{subfigure}[b]{0.3\textwidth}
         \centering
         \includegraphics[width=\textwidth]{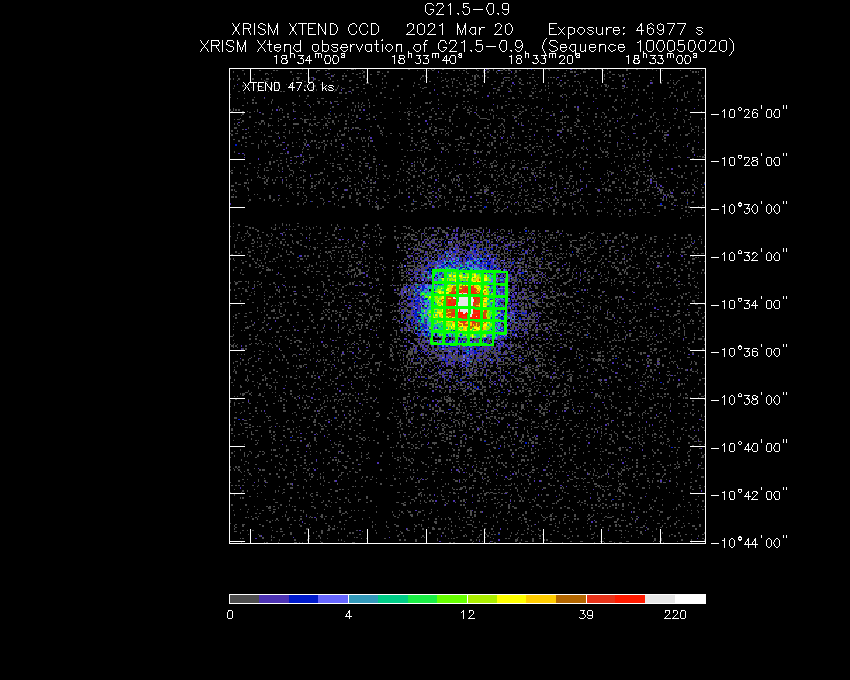}
         \caption{XRISM/Xtend image}
         \label{fig:xtdimg}
     \end{subfigure}
     \hfill
     \begin{subfigure}[b]{0.3\textwidth}
         \centering
         \includegraphics[width=\textwidth]{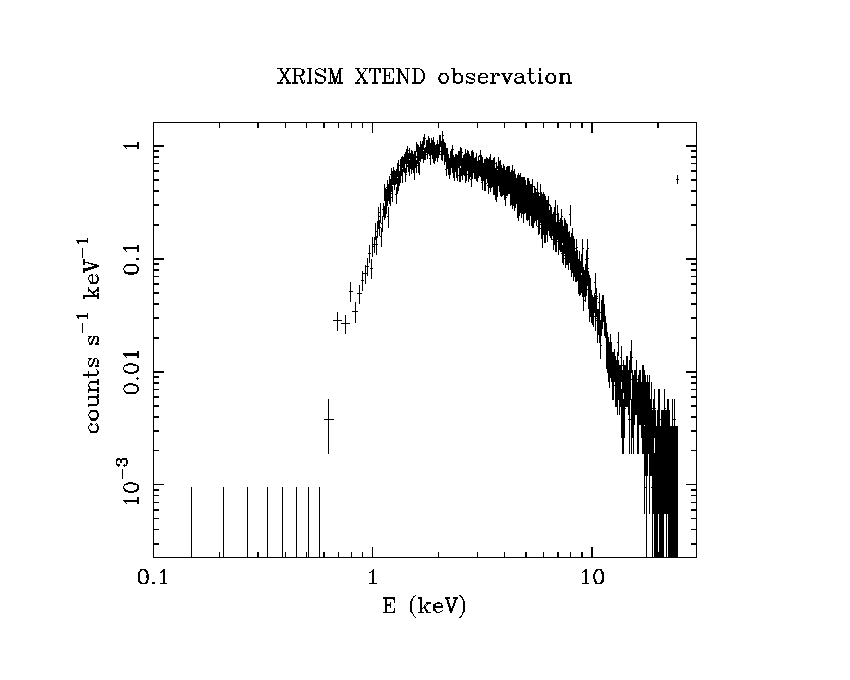}
         \caption{XRISM/Xtend spectrum}
         \label{fig:xtdspec}
     \end{subfigure}
     \hfill
     \begin{subfigure}[b]{0.3\textwidth}
         \centering
         \includegraphics[width=\textwidth]{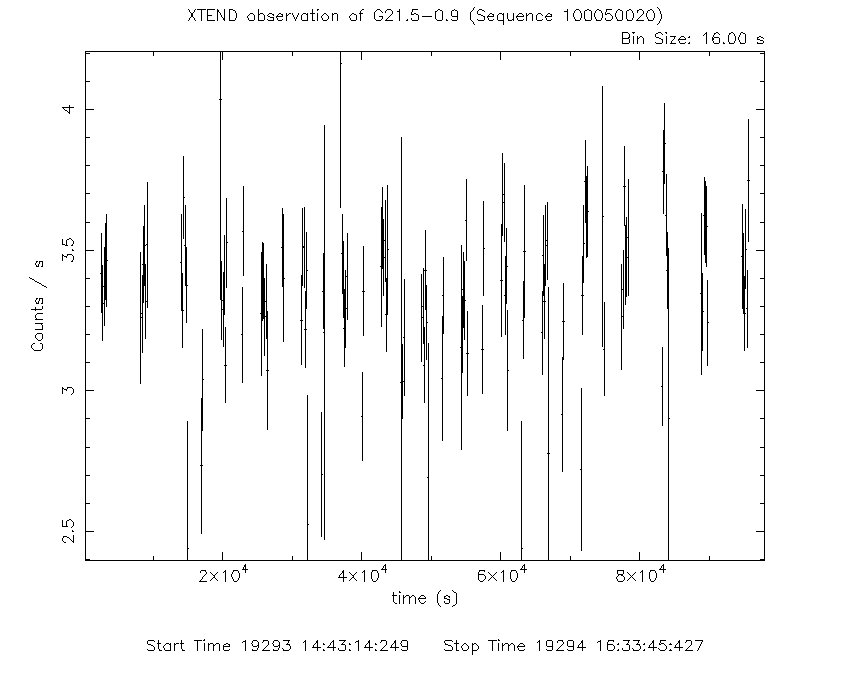}
         \caption{XRISM/Xtend light curve}
         \label{fig:xtdlc}
     \end{subfigure}
        \caption{XRISM/Xtend pipeline products from the the Hitomi observation 100050020, converted to be compatible with XRISM software, see \S\ref{sec:minimission} for more details.}
        \label{fig:prodxtd}
\end{figure}

The pipeline processing script not only produces processed data, but also preview products for each XRISM instrument.  The SDC has improved and expanded the generations of preview products since Hitomi.  These preview products include spectra, light curves, and images, for both Resolve and Xtend instruments, plotted at preset wavelength ranges.  The purpose of these preview products is for scientists to be able to quickly browse an observation to determine if any features of great interest exist.  The pipeline products will be available as thumbnails when browsing XRISM data in the HEASARC archive.  Fig.~\ref{fig:prodrsl} displays an example of an image, spectrum, and light curve for XRISM/Resolve, while Fig.~\ref{fig:prodxtd} displays the same for XRISM/Xtend.

\subsection{Coordination with JAXA/ISAS}
There has been a considerable pre-launch, bi-continental, and collaborative effort to test the communication of the PL with the PPL.  This effort has involved producing a working PL, a working PPL, and connecting the two with email communication, xDTS, DAS, and team member communications.  This effort has involved the cooperation of both GSFC pipeline operators and ISAS PPL operators.

The process of testing the data flow from Japan to the US and back to Japan, including PPL and PL processing, is a procedure we've termed an ``end-to-end test''.  First at ISAS, the PPL is executed and FFFs are produced for a test sequence.  Then there is a test of communications, which notifies GSFC of data ready to be transmitted from ISAS.  The data readiness automatically triggers the PL to start at GSFC.  The sequence is then processed through the PL and the data is transferred back to the archive at ISAS.  Not all functionality of the end-to-end test is fully functional at this point pre-launch, but there is enough functionality to transfer data and validate both PPL and PL output, which both teams have conducted with each end-to-end test.

Validation efforts are completed with each SDC internal build release of the pre-launch XRISM-specific software tasks and with each end-to-end test.  We have produced test data sets, discussed in \S\ref{sec:minimission}, with which we can (a) validate the output from the PPL and the interface between the PPL and PL, (b) validate the output from the PL, and (c) ensure the communication and transfer of data is working properly.  We have recently been able to test the archiving function of the pipeline, compressing and encrypting necessary data, and distributing data into sub-directories in the archives.  We are beginning to conduct the interfacing with the archives themselves, both DARTS and HEASARC.  The next steps are to test database creation and functionality and full archive testing.

\section{Mini Mission Data} \label{sec:minimission}
The mini mission data are a set of sequences used for testing the PL, user reprocessing scripts, and individual XRISM-specific software tasks.  The main overall purpose of the mini mission data is: (1) for internal (SDC only) testing of software tasks and pipeline, (2) to set up an interface between the PPL team and the SDC for what the PL is expecting the FFFs to contain and for correct formatting, and (3) provide test data for the broader XRISM team to use for testing of software, learning how to use the XRISM software, and for comparison to ground system processing.  The mini mission data consists of four types of data sets: (1) archival Hitomi data converted to XRISM software-compatible data, (2) pre-pipeline output from ISAS, (3) Resolve instrument ground testing and calibration data, and (4) Xtend instrument ground testing and calibration data.  Each type of data set is generally used for a different testing purpose.

The archival Hitomi data converted to XRISM software-compatible data was the easiest and most accessible, since all Hitomi data is publicly available on HEASARC.  Since the Hitomi/SXS instrument is nearly identical to XRISM/Resolve and Hitomi/SXI is nearly identical to XRISM/Xtend, the conversion from Hitomi to XRISM data was straightforward.  Most of the conversion involved shifting times and dates five years ahead to account for the new mission epoch (2014 for Hitomi changed to 2019 for XRISM) and updating instrument-specific keywords and file names to match those of XRISM instead of Hitomi.  We used five sequences from the archival Hitomi data, both the pipeline-processed data that was in the public archive, and the FFFs from the PPL, which were in the SDC internal archive (obtained from the deep archive).  This provided test data to use both on the PL and on XRISM-specific tasks and user pipeline reprocessing scripts.  These data were also updated in accordance with the XRISM Science FITS ground system interface control document.  By creating this data set, we were able to use this as an interface to the PPL team to provide them with the formatting and contents of what the PL is expecting.

As previously discussed, the XRISM SDC has been conducting end-to-end tests with ISAS.  The ISAS and GSFC teams have been working together and increasing the complexity of data transferred over the past couple of years.  The first tests used Hitomi data as a simple test of the transfer of data and allowed us to perfect the interfacing of the PL and PPL systems.  We then progressed to an early version of the Hitomi-converted to XRISM software-compatible mini mission data, which we artificially placed on PPL outgoing servers, transferred to the PL servers, and processed the data through the PL.  We are currently using data output from the current pre-launch state of the PPL, transferred to the PL and processed through the PL, then sent back to ISAS.  The SDC has been validating the output PPL data and checking for compliance with the XRISM Science FITS and archive interface documents, and the interfacing and communication of the PPL with the PL.

The third set of mini mission data is the Resolve ground testing and calibration data.  This is data we have received from the Resolve instrument team, which has been used for ground calibration and testing.  The Resolve ground data is best suited for testing our user pipeline reprocessing scripts; however it is necessary to add in supplementary files and make minor modifications to the existing files in order to create a data set that can be run through the reprocessing scripts.  Since this data is taken on the ground, it does not have corresponding spacecraft auxiliary files such as attitude and orbit, so these either have to be artificially created, or use Hitomi-converted data, with times updated to match the ground data timing.  The data provided to the SDC by the Resolve instrument team is also a good check of updates, additions, and changes of keywords and such that have been discussed and agreed upon by the two teams.  This data aids in communication and interfacing between the Resolve instrument team and the SDC.  

The fourth set of mini mission data is the Xtend ground calibration and testing data.  Similar to the Resolve ground data, this data was provided to the SDC by the Xtend instrument team following ground calibration and testing.  There are also similar caveats to the Xtend ground data including having to supplement the data with converted auxiliary files.  Similar to the Resolve ground data, we have tried to keep the Xtend data as intact as possible, updating times of converted auxiliary files to match those of the ground data.  The Xtend ground data is used to interface with the Xtend team and to aid in the testing and validation of Xtend-specific software tasks.

The conversion of the Hitomi archival data sets into XRISM software-compatible data was achieved by using a number of FTOOLS and FITS manipulation functions linked together in Python scripts.  The PPL and ground data conversions were conducted in a similar manner as the archival data, but fewer modifications were necessary.  We used tools to add columns, update header keyword values and comments, add in new keywords, delete keywords, update timing keywords and columns, and so forth.  Internal documentation has been produced to record the changes performed on the original data and input data, output data, and the scripts produced have been made available to the team.

\section{Unit Test Framework} \label{sec:unittests}
We have been working to greatly improve the unit test framework that was first set up by Hitomi.  The automated unit test execution is performed by a HEASoft task called \textsf{aht}.  The \textsf{aht} task can create a unit test, and execute a unit test and report a pass or fail.  We have written a wrapper script that can run all or a specified set of tests and produce a succinct output of the outcome of all tests that were executed.  The major issue for our team with the \textsf{aht} framework is the multiple copies of data \textsf{aht} creates when creating and validating unit tests.  Some of the XRISM data is quite large in size and the tests use shared files from the CalDB that are repeated in several tests.  Additionally, some tests have large input/output files that have multiple copies stored for a single test.  These large files caused our git repository to become bloated with extra data.  We are currently investigating an addition to the \textsf{aht} framework that would allow for shared directories of read-only input files, instead of each unit test keeping a separate directory of input files.

\begin{figure}[h!]
    \centering
    \includegraphics[width=\textwidth]{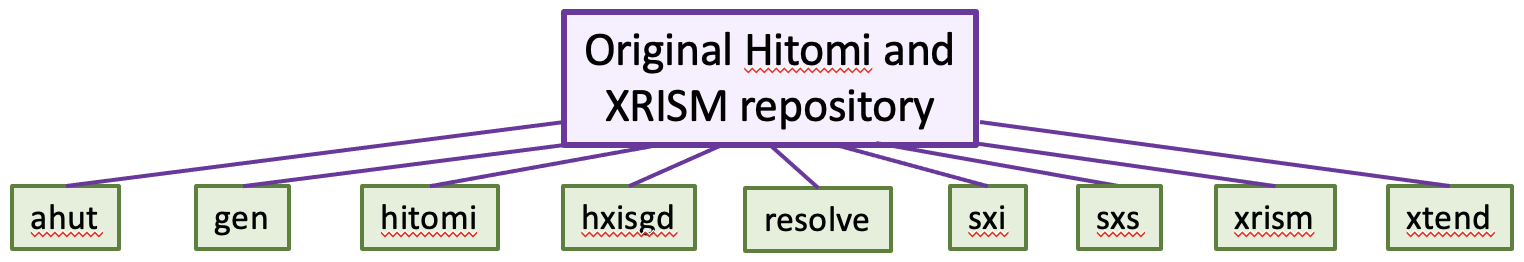}
    \caption{Original unit test repository setup, where all unit tests were contained in a single git repository.}
    \label{fig:origrepo}
\end{figure}

\begin{figure}[h!]
    \centering
    \includegraphics[width=\textwidth]{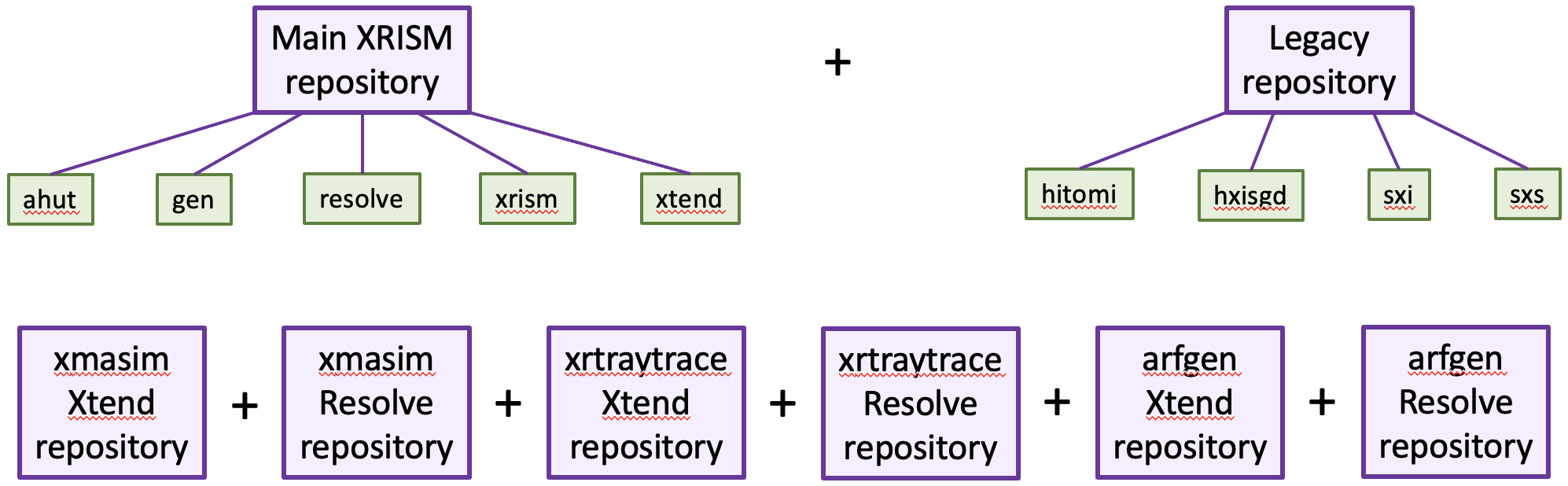}
    \caption{New unit test multiple git repository setup. The tasks that have their own repositories have large data sets and therefore needed to be separated from the main XRISM repository.}
    \label{fig:newrepo}
\end{figure}

Originally, all unit tests for all tasks under the XRISM SDC purview, with the exception of information only tasks, were stored in a single, local (not GitHub) git\footnote{\url{https://git-scm.com/}} repository.  This repository contained sub-directories generally dividing groups of tests based on some similarity, i.e., instrument-specific tasks, as can be seen in Fig.~\ref{fig:origrepo}.  Some of our unit tests contain large input/output data sets and therefore bloat the git repository.  We decided to separate the unit tests into multiple git repositories, to keep the size down and make the repositories easier for transfer to developer machines and for updating and running the full set of unit tests.  The new structure of our git repositories for all XRISM and Hitomi unit tests is displayed in Fig.~\ref{fig:newrepo}.

\begin{figure}[h!]
    \centering
    \includegraphics[width=0.75\textwidth]{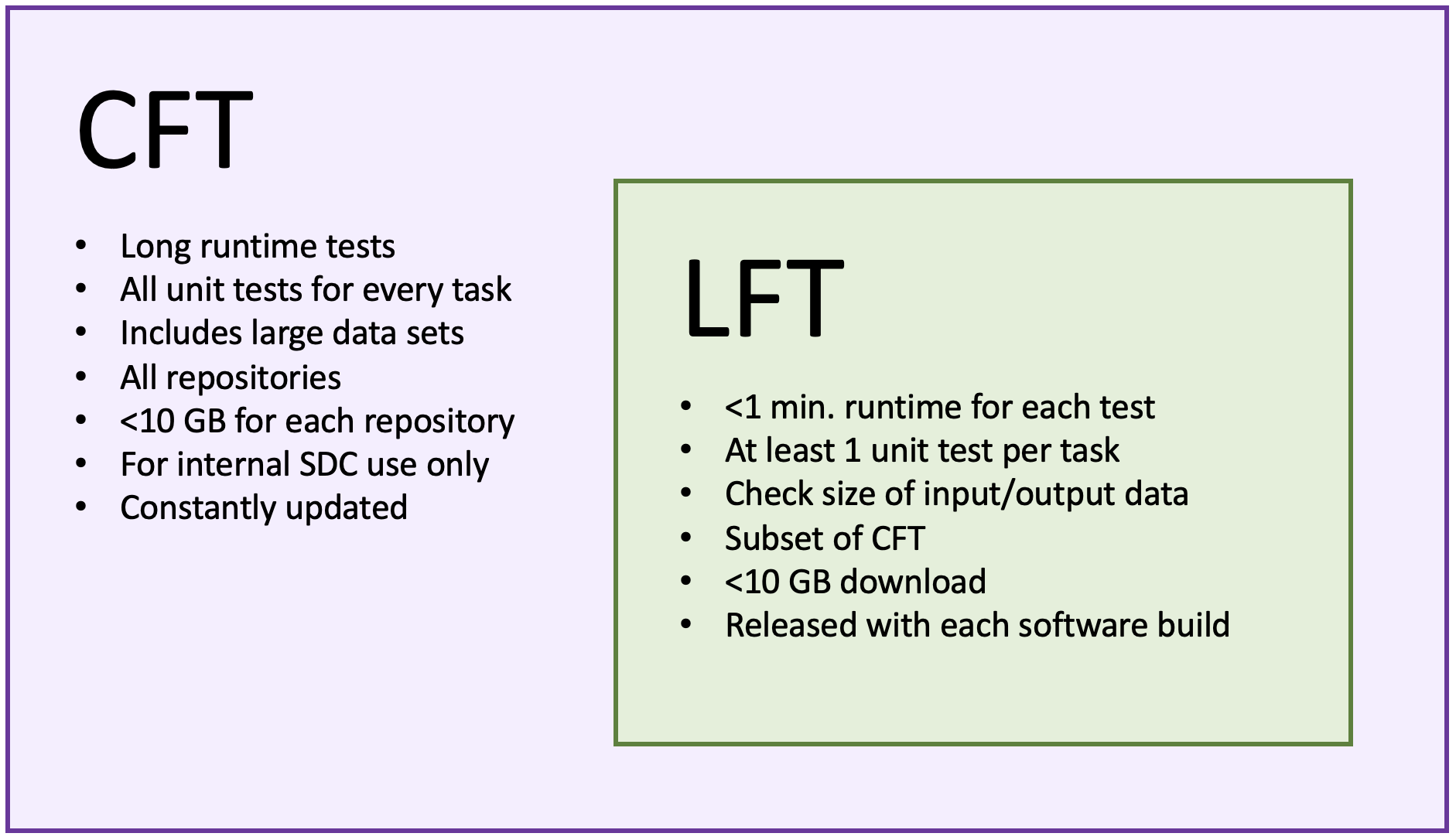}
    \caption{Quick comparison of CFT vs. LFT to highlight the major differences.}
    \label{fig:cftlft}
\end{figure}

Additionally, we decided to set up a new framework of unit test sets.  We created a set of limited functional tests (LFT) and a set of comprehensive functional tests (CFT).  The CFT is the full set of unit tests for all 62 tasks and includes the full version of all of the git repositories.  The CFT is only for internal use, mostly for preparing for build releases of our software and regression testing.  The CFT includes very long-running and large data set tests, which are more difficult to provide to a general user.  The LFT will be deployed with our XRISM software to aid in user support.  The LFT is a light-weight subset of the CFT, which still contains tests for all 62 tasks, but only select tests for each task.  A general comparison of the CFT and LFT are shown in Fig.~\ref{fig:cftlft}.

The original unit tests were obtained from the Hitomi mission.  Most of the mission-specific and instrument-specific unit tests were copied and converted to XRISM unit tests.  For example, unit tests from the Hitomi/SXS \textsf{sxspha2pi} task were copied and recreated for the corresponding XRISM/Resolve \textsf{rslpha2pi} task.  Similarly for general Hitomi \textsf{ah} tasks like \textsf{ahmkehk}, the unit tests were copied and converted to unit tests for the general XRISM \textsf{xamkehk} task.  The Hitomi unit tests have been retained for regression testing, particularly after a XRISM task has been updated to deviate from the original Hitomi task.  Nearly all of the XRISM unit tests have recently been reviewed, validated, updated, and in some cases replaced; the validation process is ongoing.  New unit tests have been added with formal development of XRISM tasks, or introduction of new tasks.  Each task is currently undergoing, or has undergone, a careful review to ensure that the set of unit tests for the particular task covers all functionality as laid out in the requirements of the help file (see \S\ref{sec:helpfiles} for more on help files).

\section{Documentation} \label{sec:documentation}
Documentation is extremely important in software development, both for the developers, and for the users.  We strive to provide documentation with ease of readability and understanding, coherence, and consistency for developers, XRISM team members, and general users.  In this section, we describe the three types of documentation the XRISM SDC has employed to try to achieve the aforementioned goal.  We discuss user documentation of our software tools (\S\ref{sec:helpfiles}), documentation of the code itself (\S\ref{sec:doxygen}), and internal documentation for interfacing and recording necessary information for team use (\S\ref{sec:intdocs}).

\subsection{User Documentation: Help Files} \label{sec:helpfiles}
Each task within the XRISM software suite has a corresponding help file.  Help files are a construct of all HEASoft tasks, which provide a general description of the task, general usage of the task, examples of how to run the task, and lays out all the input/output parameters of the task.  Help files can be accessed on the HEASoft website, as well as on the command line (after the HEASoft environment has been initialized), using the command \texttt{fhelp taskname}, where \texttt{taskname} is the name of the task.  The parameters are described in detail in the help files, with options for parameters that have a limited number of options, and provides functionality for when a parameter is set to a specific value.  The help file also displays the parameters that are required for execution of the task, and those that are optional.

The XRISM SDC conducted a thorough review of all the help files of the tasks under its purview, and has updated the files for understanding and accuracy.  The task descriptions were carefully reviewed, and in some cases rewritten, for ease of readability and understanding.  The examples were updated with XRISM data input/output file names, to allow for cohesiveness of the XRISM software and example data.  We reviewed the help files for consistency in formatting and coherence in content as well.  Since the help files essentially lay out the functionality of the task, we used these as a guide for the results that a particular task should achieve, given particular input parameters and/or input data.  This provided us with a baseline for how to test each task with unit tests. As discussed previously in \S\ref{sec:unittests}, the help files were used as a guide for creating unit tests for each task, to ensure that essential functionality was properly tested, and that the task performed as expected, according to the prescribed functionality.

\subsection{Code Documentation with Doxygen} \label{sec:doxygen}
The previous Hitomi code included Doxygen \cite{doxygen} markup, but was not consistent enough or fully implemented throughout all tasks.  For XRISM, we have implemented consistent Doxygen markup in all code files for each task; this includes *.c, *.cxx, *.h, *.pm, *.pl, and *.dox files.  We performed an overhaul of the Doxygen documentation within the code modules to create consistent and comprehensive documentation that would be presented in a coherent manner.  The header at the beginning of each code file consists of the following documentation keys: file name, brief description, author, date of last code update, version number, definition of code groups, longer description, source files, subroutines (for Perl only), tool dependencies, library dependencies, author history, and modification history.  Doxygen markup is also used throughout the code to document classes, functions, structures, variables, etc.  The C and C$^{++}$ code is automatically assembled by Doxygen, but the Perl integration is still a work in progress, so we have created a general workaround by duplicating the Doxygen header information from any Perl modules or scripts into a corresponding .dox file.  The .dox files are picked up by Doxygen in a similar manner to the C/C$^{++}$ code files, however the full functionality is not equivalent for the Perl scripts.

\begin{figure}[p!]
     \centering
     \begin{subfigure}[b]{0.42\textwidth}
         \centering
         \includegraphics[width=\textwidth]{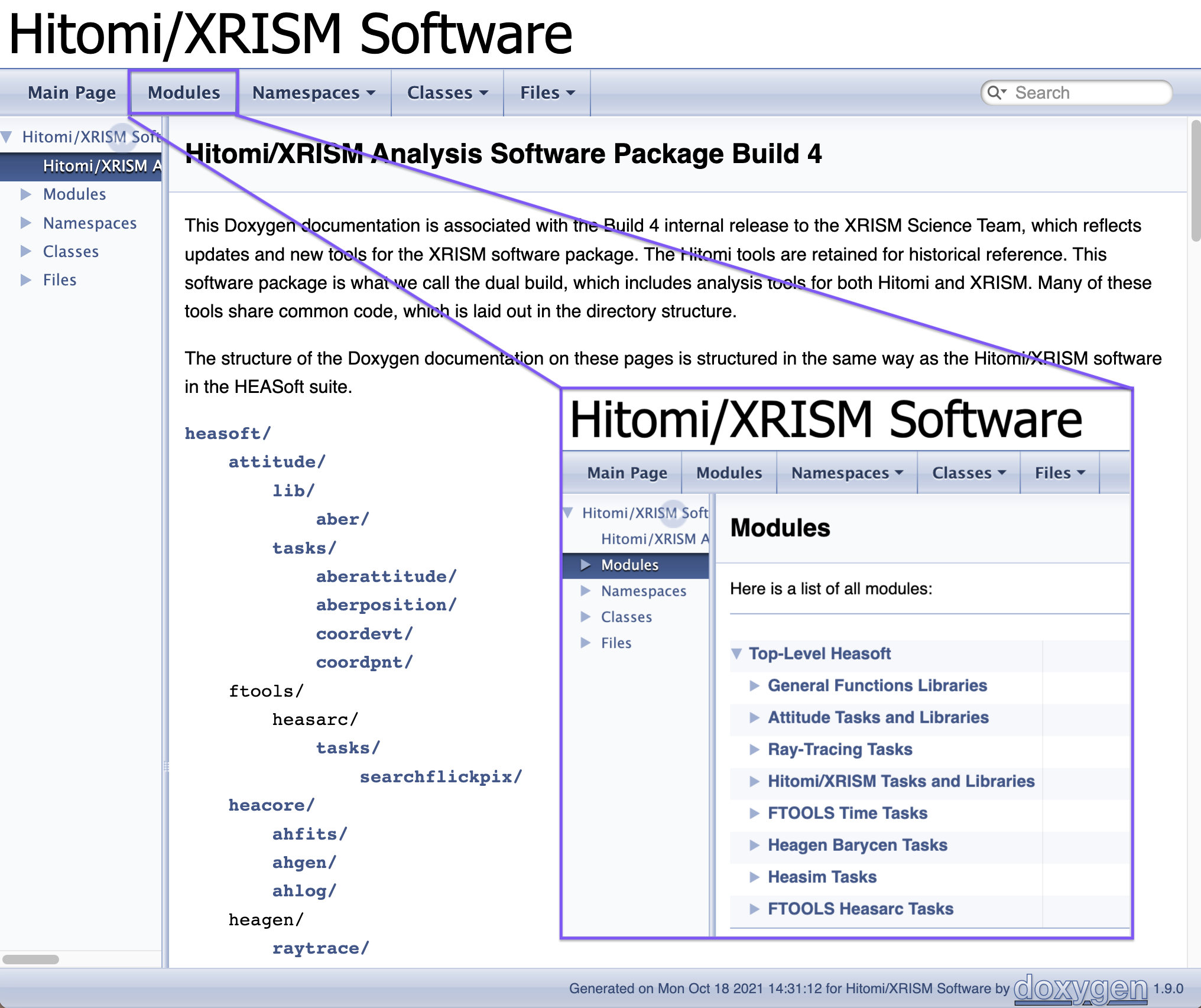}
         \caption{The main page of the Doxygen HTML output. The inset shows a snapshot of the page when the Modules tab is clicked on.}
         \label{fig:doxmodules}
     \end{subfigure}
     \hfill
     \centering
     \begin{subfigure}[b]{0.42\textwidth}
         \centering
         \includegraphics[width=\textwidth]{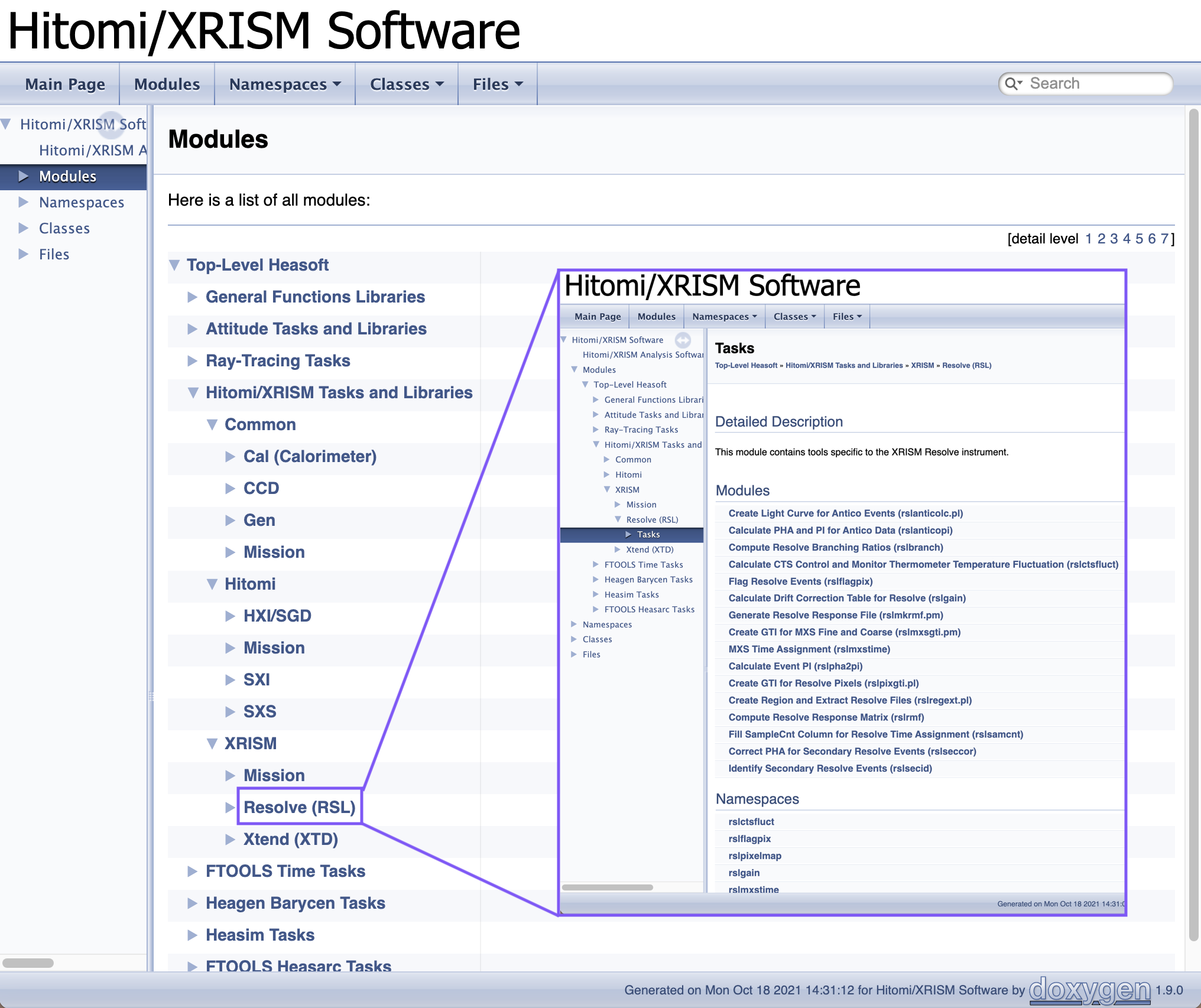}
         \caption{Here is the Modules page. The inset shows the page when the Resolve (RSL) module is clicked on.}
         \label{fig:doxrslmodules}
     \end{subfigure}
     \hfill
     \centering
     \begin{subfigure}[b]{0.42\textwidth}
         \centering
         \includegraphics[width=\textwidth]{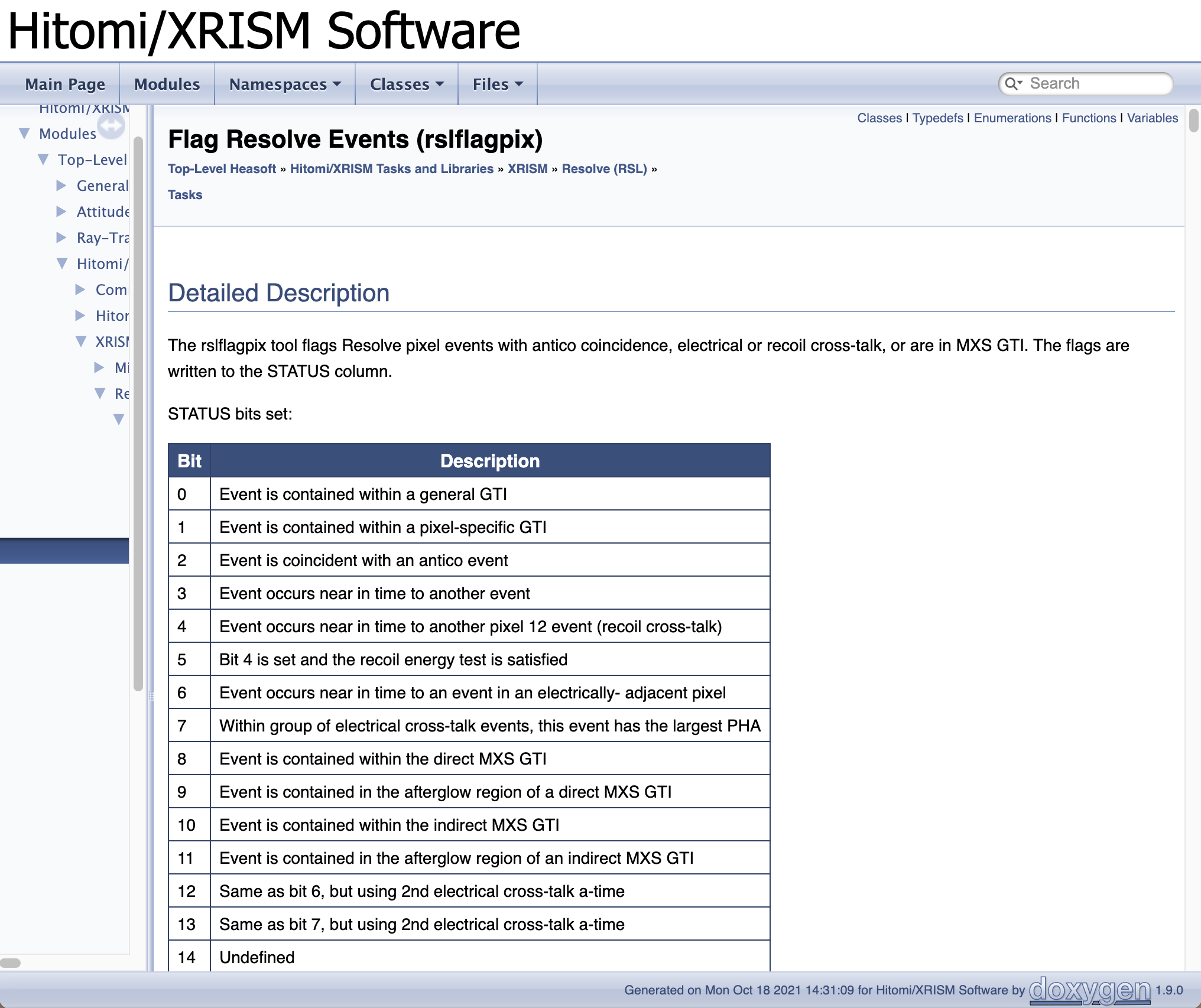}
         \caption{Example of the description of a task when it is clicked upon from the Resolve modules page.}
         \label{fig:rslflagmain}
     \end{subfigure}
     \hfill
     \begin{subfigure}[b]{0.42\textwidth}
         \centering
         \includegraphics[width=\textwidth]{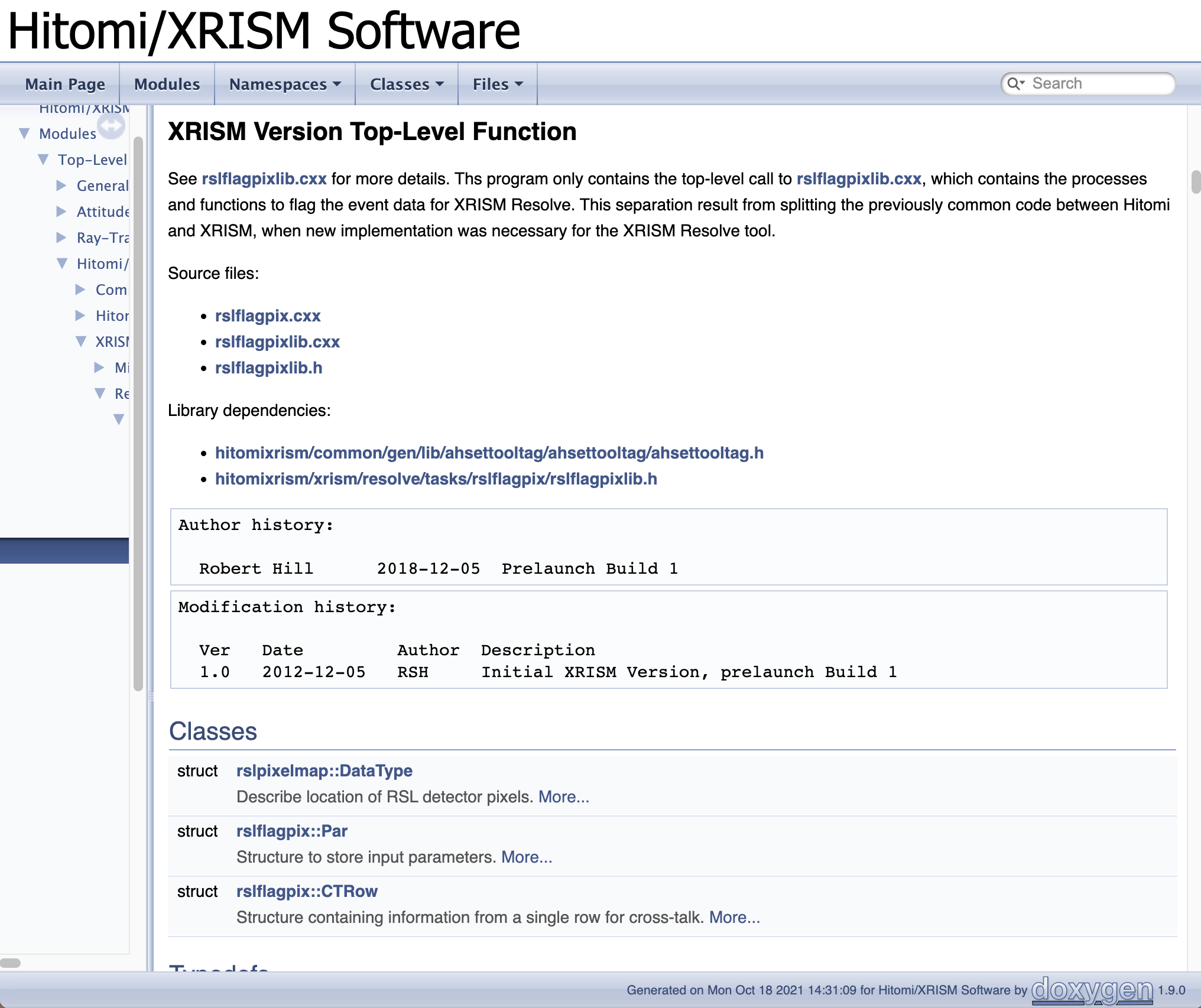}
         \caption{Example of other sections of the header.}
         \label{fig:rslflaghead}
     \end{subfigure}
     \hfill
     \begin{subfigure}[b]{0.42\textwidth}
         \centering
         \includegraphics[width=\textwidth]{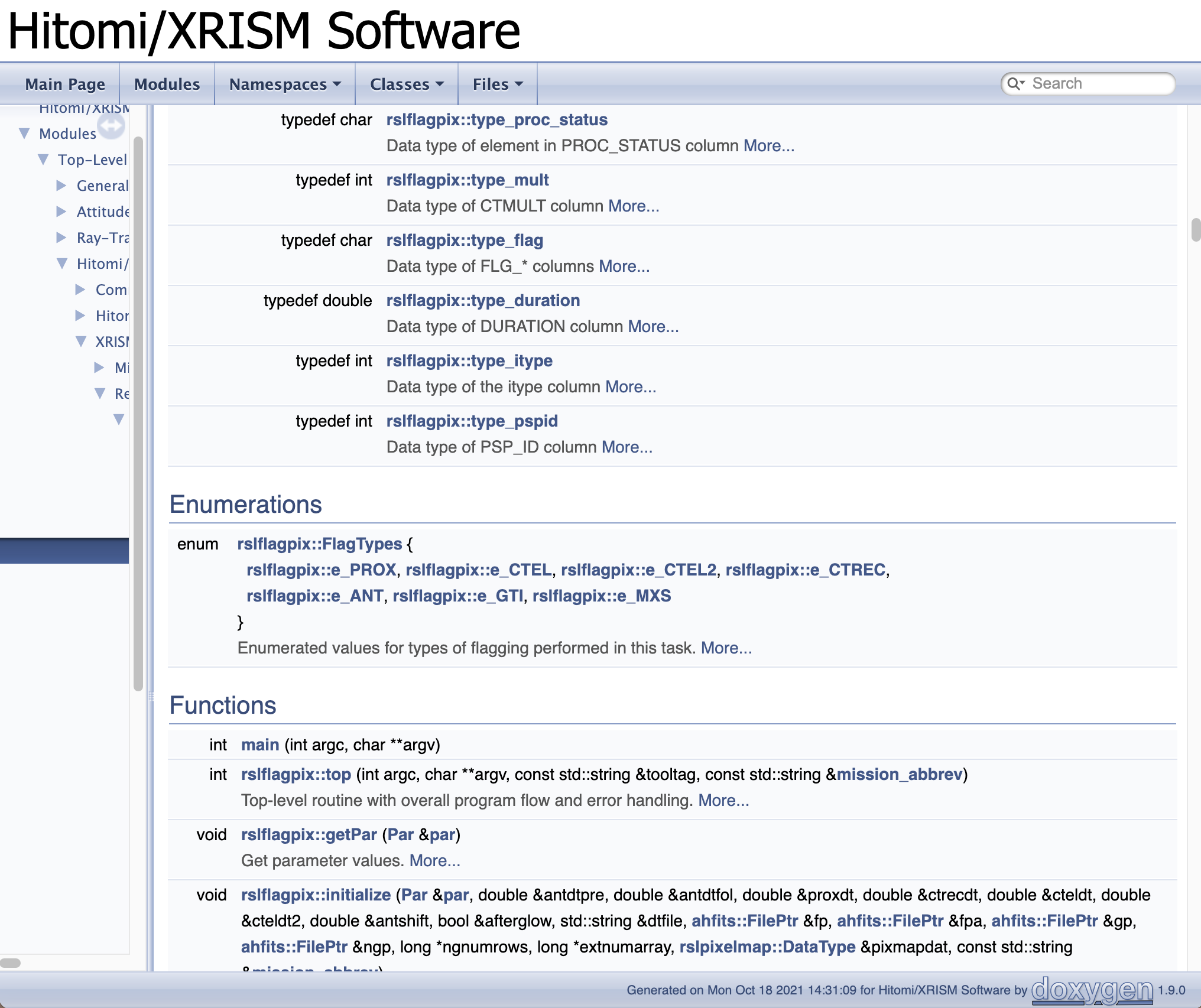}
         \caption{Example of the classes of a task.}
         \label{fig:rslflagclass}
     \end{subfigure}
     \hfill
     \begin{subfigure}[b]{0.42\textwidth}
         \centering
         \includegraphics[width=\textwidth]{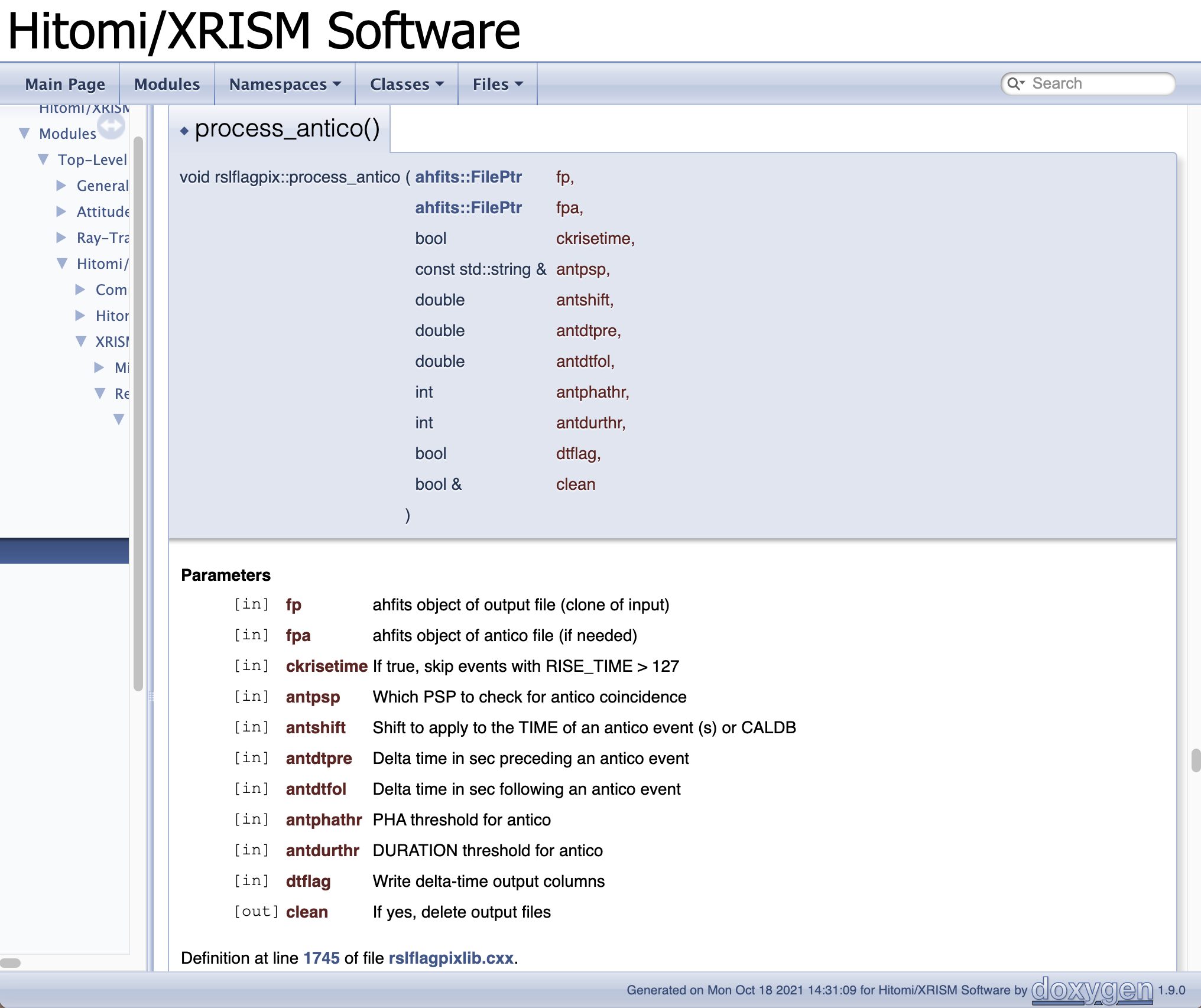}
         \caption{Example of a function declaration.}
         \label{fig:rslflagfunc}
     \end{subfigure}
        \caption{Here we show an example of some of the components of a task in the Doxygen HTML output.}
        \label{fig:doxcomponents}
\end{figure}

We use the HTML option for output from executing Doxygen on the XRISM code files.  We packaged the HTML bundle and it has been released with pre-launch software builds.  The HTML is local to the user's machine, so it does not need to be hosted and can remain internal, at least for now, and tied to a specific software release.  The Doxygen HTML layout follows the directory structure of tasks within HEASoft, therefore making it easier for the user to follow.  The documentation is fully clickable through functions and the code itself.  The Doxygen output provides a way for users who want to delve a bit deeper into the code or the algorithms, to be able to easily navigate the code without having to slog through the full source code.  Fig.~\ref{fig:doxcomponents} displays an example of the main pages of the Doxygen HTML output as well as some screenshots of what a specific task would look like for a user.

\subsection{Internal Documentation} \label{sec:intdocs}
We use an internal document tracking and storage system to post, review, and update major documents, which can be accessed by the entire NASA/GSFC-based XRISM team.  The XRISM SDC has produced a collection of documents, including reference documents from Hitomi, and new and updated documents for XRISM referring to the pipeline, its processes, products, and interfacing between JAXA/ISAS and NASA/GSFC.  

We also produce engineering change request (ECR) documents, which can either describe a completely new tool, or modifications to an existing tool. In the latter case, the ECR provides details of the issue, background, suggested code changes to solve the issue, and proposed tests for an update or modification to the code.  These documents are written by one or more of the SDC scientists and are implemented by one or more SDC developers, depending on the magnitude of code changes.  The scientists and developers work together through implementation of the ECR modifications to the code and testing.  A reconciliation document is then produced by a developer or developers, whereby the changes line-by-line in the code are tracked, any comments about implementation are recorded, and tests are recorded with expected outcomes and actual outcomes.  We also produce SDC-only internal documentation such as documenting the mini mission data creation, unit test creation, pipeline validation, pipeline processing script processing steps, and more.  Many of these are on shared websites for easy access to collaborative editing, with hard copies generally on our shared server.  We produce these detailed documents in order to maintain our processes, enable reproducible results, document communication and decisions made, and to have a ``paper trail'' of software modifications.

\section{Improvements Since Hitomi} \label{sec:improvements}
As noted previously, the XRISM SDC acquired the legacy Hitomi software.  Over the past 4+ years, we have created a suite of software tasks that will be ready for the XRISM launch.  The suite is currently composed of 
what we call the dual build, which is to say that there exists some code that has not changed from Hitomi and is therefore common between Hitomi and XRISM, but also Hitomi and XRISM tasks that have now deviated from one another and are wholly separate.  For tasks with this ``common code'', the Hitomi and XRISM counterparts contain the exact same code and behave exactly the same, with only essential mission-specific keywords and parameters diverging. 

We still retain some common code, but many of the tasks have now been split.  Bugs found in the Hitomi code will eventually be corrected but currently, only the XRISM versions have been corrected, as a priority.

The XRISM SDC has made modifications to 12 Resolve tasks, 3 Xtend tasks, 11 mission-specific tasks, and 3 other tasks, a total of 29 out of 62 total tasks under the XRISM SDC purview.  We have also deleted some tasks, which are no longer relevant to the XRISM mission.  The Resolve and Xtend task updates have been informed by the respective instrument teams, whether there has been updates to the instrumentation itself, or the analysis of the calibration data.

We work closely with the Resolve instrument teams, both in the US and in Japan, as well as the Xtend instrument team in Japan.  We take input from scientists, the mission operations team, and the science operations team.  We have forged strong ties with these teams in order to create a communicative environment for updating software, validating data, and testing the pipeline.  We have also performed instrument-driven updates to for CalDB files.

Major tool improvements already completed include the following:
\begin{itemize}
    \item Run time and functionality improvements for our raytracing program, \textsf{xrtraytrace},  which traces X-ray photons through a telescope
    \item A first pass at run time optimization (more planned) and introduction of new components and functionality for the Resolve RMF generator
    \item Near completion of the X-ray Spectral Line IDentifier and Explorer (XSLIDE) tool, which is a quick-look spectral analysis tool for XRISM data \cite{Braun2022} (see Braun et al., these proceedings)
    \item Improvements to the user pipeline scripts, \textsf{xapipeline},  which reprocesses both Resolve and Xtend data, \textsf{rslpipeline}, which reprocesses only Resolve data, and \textsf{xtdpipeline}, which reprocesses only Xtend data
    \item Significant improvements and algorithm changes have also been performed for flagging pixels in both Resolve (\textsf{rslflagpix}) and Xtend (\textsf{xtdflagpix}), pulse invariant calculations for both Resolve (\textsf{rslpha2pi}) and Xtend (\textsf{xtdpi})
\end{itemize}

We have also performed some bug fixes, which have been found either during formal development, reported by a team member, or discovered in our intensive testing process.  Several minor modifications have also been performed for many of the XRISM tasks.

\subsection{Future Improvements}
The XRISM SDC has plans for future improvements to tasks and new tools, mostly for supporting post-pipeline data analysis and creating simulated data.  We plan to optimize the runtime in order to create a more efficient RMF generator.  We will continue to make CalDB updates and improvements with better calibration and data from the Resolve and Xtend instrument teams.  CalDB updates will continue after launch, when there are likely to be minor tweaks due to instrument performance that is different in-flight, compared to what it was during ground testing.  

We have started development of a raytracing driver tool, which will allow more flexibility and functionality for simulations that trace X-ray photons through an X-ray telescope and either of the detectors. We intend to create a point spread function (PSF) library for Resolve, which would contain pre-computed PSFs for use by scientists, and would alleviate the burden of having to execute extremely time-consuming tasks for generating the effective area and/or PSF for arbitrary satellite pointing parameters, and detector selection regions. We are also considering an RMF library, which would contain pre-computed RMFs for users, again so a user would not have to run extremely time-consuming tasks to create large RMFs.  We also have plans to make improvements to the efficiency and functionality of the auxiliary response function (ARF) generator \cite{yaqoob2018}.  A new tool will also be developed that will account for the
effect of intervening galactic dust halos on the effective area and PSF.  Many of these planned improvements will be slated for after launch, to ensure the core software and the pipeline are ready for data processing.


\bibliographystyle{spiebib} 
\bibliography{spieproc2020_xrism_sdc.bib} 

\end{document}